# UV and deep-UV ultrafast harmonic generation enhanced by resonant gold nano-pillar metasurface

**Shroddha Mukhopadhyay** ,[1,*] **Yuxuan Chen**,[2] **Hooman Barati Sedeh**,[2] **Natalia M. Litchinitser**,[2] **Michael Scalora**,[3] **Jose Trull** [1], **Maria Antonietta Vincenti**,[4] **Crina Cojocaru**[1]

[1]*Department of Physics – Universitat Politècnica de Catalunya, Rambla Sant Nebridi 22, 08222 Terrassa (Barcelona), Spain*
[2]*Department of Electrical and Computer Engineering, Duke University, Durham, NC 27708, US*
[3]*Institute of Photonics and Quantum Sciences, Heriot-Watt University, Edinburgh, EH14 4AS, UK*
[4]*Department of Information Engineering – University of Brescia, Via Branze 38, 25123 Brescia, Italy*
*shroddha.mukhopadhyay@upc.edu

**Abstract**

We report experimental observations and full-vector numerical simulations of second and third harmonic generation from a two-dimensional periodic gold nano-pillars metasurface, for efficient ultrafast light conversion from near infrared light to ultraviolet and deep-ultraviolet spectral ranges, respectively. The metasurface is designed to support a characteristic plasmonic resonance, which is tunable across the near-infrared spectral range, with its resonance wavelength shifting as a function of the pump beam angle of incidence. Under resonant conditions, the nanopillar structure strongly confines the electromagnetic field near the metal surface, producing substantial field enhancement at the pillar edges and thereby boosting nonlinear optical processes. Compared with a flat gold mirror, the nanostructured metal exhibits a pronounced enhancement in nonlinear frequency conversion from the visible to the ultraviolet spectral range, with conversion efficiencies of both second- and third-harmonic generation increased by more than two orders of magnitude. Our results are validated by full-wave theoretical modeling of linear and nonlinear light-matter interaction in nanoscale metallic structures. We demonstrate that plasmonic structures comprising periodically arranged square pillars can be engineered to achieve enhanced second- and third-harmonic generation at wavelengths that can be tailored through the structural design, without requiring complex geometries or hybrid structures that pose fabrication challenges. The close agreement between theoretical predictions and experimental measurements further advances our understanding of surface-mediated nonlinear optical processes in metals and highlights the potential for further exploration of nonlinear optical phenomena in metallic nanostructures and new ultrafast light sources in the UV and deep-UV.

## 1. Introduction

Metasurfaces, planar arrays of subwavelength resonant elements, have emerged as a powerful platform for manipulating light–matter interactions far beyond the limits of conventional optics. By tailoring localized surface plasmon resonances (LSPRs), surface lattice resonances (SLRs), and hybrid multiresonant modes, plasmonic metasurfaces can concentrate electromagnetic (EM) energy into deeply subwavelength volumes, enabling dramatic enhancement of both linear and nonlinear optical processes. Over the past decade, a wide variety of architectures, including nanopillars, nanopatches, nanotrenches, and plasmonic gratings, have been developed to exploit these effects for sensing, spectroscopy, and nonlinear frequency conversion [5-9, 19, 23, 24, 29, 30, 46].

Nonlinear frequency conversion in metals has been central to nonlinear optics since its inception, driven by the large third-order susceptibilities of plasmonic materials despite their intrinsic absorption in the visible (VIS) and near-infrared (NIR) spectral ranges [1, 4]. Early investigations focused primarily on planar films and one-dimensional (1D) periodic structures such as gratings, whose symmetry restricts device functionality to a single polarization. Advances in nanofabrication, particularly electron-beam lithography, now enable the realization of fully two-dimensional (2D) periodic metal architectures with precise control over periodicity, aspect ratio, and feature size. Corrugated metallic nanostructures with surface reliefs on the order of 10–100 nm can strongly reshape the spatial distribution of incident EM fields and provide substantially enhanced field confinement compared with planar metallic films. Purpose-built metasurfaces exploit plasmonic resonances - collective oscillations of free electrons that couple strongly to incident light at specific wavelengths [5–7] - to concentrate EM fields near metal surfaces. This confinement depends sensitively on the geometry of the nanostructure, including its shape, size, and sharpness, as well as on the surrounding dielectric environment, and can enhance light–matter interactions by orders of magnitude. The resulting near-field intensification drives strong surface nonlinearities [7-13], enabling efficient harmonic generation at relatively low pump intensities. Surface-plasmon-enhanced second-harmonic generation (SHG) has been investigated across several grating platforms in gold [8 - 14] and silver [15 - 17], beginning with the perturbative electrodynamic treatment of metallic gratings by Agarwal and Jha in 1982 [18]. These capabilities are particularly relevant for emerging applications in compact UV sources, deep-UV photonics, and integrated nonlinear optical devices, for which planar metallic films alone provide limited field confinement and nonlinear conversion efficiency. Prior studies of SHG and third harmonic generation (THG) from ultrathin gold layers with thicknesses of 20 – 70 nm revealed intrinsic nonlinearities [3], but the achievable conversion efficiencies remained constrained by the short interaction lengths and weak field confinement inherent to planar geometries.

Advances in high-precision nanofabrication have enabled plasmonic structures that combine intrinsic nonlinear optical responses with strong local enhancement of incident EM fields. Among these, periodic gold nanopatch and nanopillar arrays have emerged as versatile platforms for linear, nonlinear, and sensing applications. Early studies by Grande and co-workers showed that finite and periodic gold nanopatch arrays support localized and collective plasmonic modes, enabling pronounced surface-enhanced Raman scattering and nonlinear signal enhancement through multiresonant field localization [19–21]. Subsequent work extended these concepts to nonlinear SERS and hybrid configurations, including doubly resonant nanopatch arrays coupled to graphene [22]. Related gold nanostructures have demonstrated highly sensitive localized surface plasmon resonance responses [23], near-perfect light trapping in two-dimensional nanotrench gratings [24], and spectral engineering through surface lattice resonances and Rayleigh-anomaly-mediated coupling [25,26]. Such collective resonances have also been exploited to enhance SHG and THG [27, 28]. Beyond nonlinear optics, periodic plasmonic arrays have enabled applications ranging from tunable structural color [29] to biomolecular detection [30], while theoretical studies have emphasized the importance of modal overlap and dispersion in determining nonlinear conversion efficiencies [31,32]. Gold nanopillar structures have additionally been explored for sensing [33], nonlinear spectroscopy [34, 35], and advanced nanofabrication approaches [36].

Despite this progress, the nonlinear optical response of pure gold nanopillar arrays remains comparatively underexplored. In particular, direct experimental studies of structures specifically designed for both SHG and THG remain limited, with much of the existing literature focused on Raman enhancement or linear LSPR-based sensing. The interplay between collective lattice resonances and the intrinsic nonlinear response of the nanopillars has also not been systematically characterized, and nonlinear characterization under ultrafast excitation and oblique incidence remains comparatively scarce. These gaps motivate the present work, in which we investigate a 2D periodic array of gold nanopillars for efficient frequency up-conversion from the near-infrared to the ultraviolet and deep-ultraviolet through SHG and THG, respectively, under ultrafast femtosecond excitation. By exploiting angle-dependent plasmonic resonances, we demonstrate two order of magnitude enhanced nonlinear light–matter interactions. The experimental observations are supported by full-wave three-dimensional (3D) finite-element simulations of the linear and nonlinear optical response. In this context, the periodic gold nanopillar metasurface provides a relatively simple platform for elucidating the physical mechanisms governing nonlinear light–matter interactions, while the same underlying principles can be extended to more complex geometries and material systems, opening new opportunities for the design of advanced nonlinear photonic devices.

## 2. Sample design and theoretical investigation

The metasurface, schematically shown in Fig. 1(a), consists of a periodic arrangement of square pillars in a square lattice, patterned on the surface of a 150 nm thick gold layer deposited on a silicon substrate. The linear spectral response of the metasurface in reflection was calculated using the Finite Element Method (FEM). 3D simulations were performed with COMSOL Multiphysics to obtain full vectorial solution of Maxwell's equation of EM beam propagation, and the spatial distributions of the electric and magnetic fields. Fig. 1(b) is a schematic representation of the computational domain corresponding to a single unit cell.. Periodic boundary conditions (PBCs) were applied along the $x$ and $y$ directions to account for the periodicity of the structure, while perfectly matched layers (PMLs) were implemented along the $z$ direction, corresponding to the propagation direction of the incident pump beam. In periodic metallic nanostructures, the lattice provides additional in-plane momentum that can facilitate coupling between the incident field and collective plasmonic or grating-mediated resonances. As a result, the spectral position of these resonances is governed not only by the local geometry and periodicity of the nanostructure, but also by the angle of incidence.

Following a detailed optimization of the geometrical parameters, the structure was defined by: p=605 nm, w=185 nm and h=50 nm, where p is the period, w is the pillar width, and h is the pillar height. The calculated reflectance map as a function of incident angle and wavelength is shown in Fig. 1(c). The plasmonic resonance is designed to lie within the near-infrared spectral range and is brought into spectral overlap with the ultrafast pump by adjusting the angle of incidence. The resonance shifts from 754 nm to 1100 nm as the angle of the incident fundamental beam increases from 12° to 22°. For an incident angle of 17°, the resonance occurs near 800 nm, as shown in Fig. 1(d). The resonance exhibits a narrow linewidth of ≈ 7 nm, corresponding to a quality factor ($Q$) of ≈ 114.3. Under these resonant conditions, the EM field is strongly localized near the nano-pillar edges and corners, resulting in intense near-field hotspots with a field amplitude enhancement factor of ≈ 40, as illustrated in Figs. 1(e) and 1(f). This resonant near-field confinement increases the local pump-field intensity, thereby promoting pump absorption and strengthening the nonlinear optical interactions, playing a central role in enhancing the SHG and THG response of the metasurface.

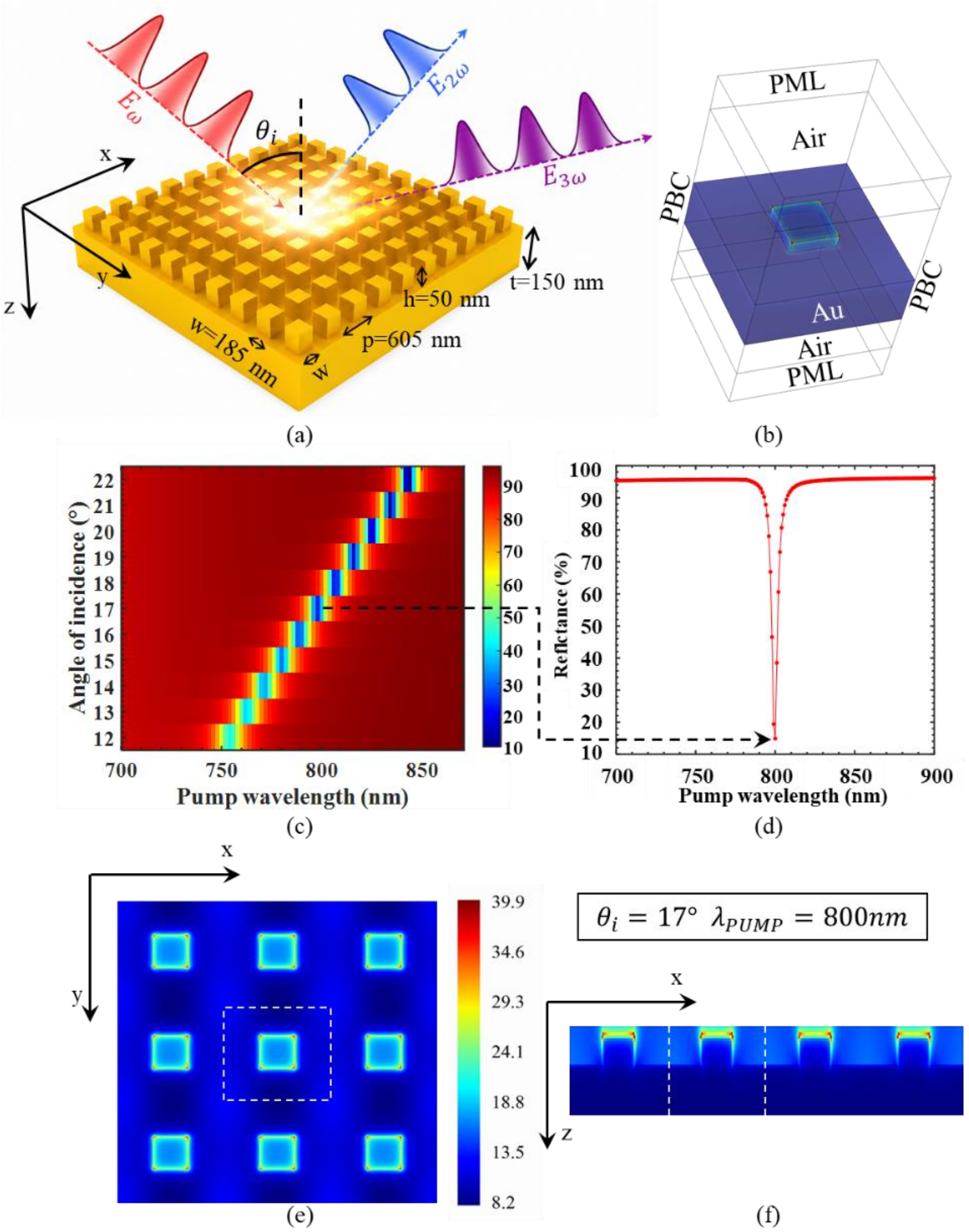

**Figure 1. (a)** Schematic representation of the gold nano-pillars (w =185 nm, h =50 nm, p=605 nm, and t = 150 nm). **(b)** Schematic of one period of the periodic gold nano pillars as modeled in COMSOL. The blue volume is gold; the corresponding boundary conditions in x, y and z directions are shown **(c)** Metasurface reflectance map for TM/TE polarized pump as a function of incident angle and wavelength. The reflectance is normalized to incident electric field intensity (representing 100%), with blue color representing low reflectance and red color representing high reflectance. **(d)** Reflectance as a function of incident pump wavelength for angle of incidence fixed at $\theta_i = 17°$. The reflectance is normalized to the incident electric field intensity (representing 100%). **(e)** Top view and **(f)** side view of electric field spatial distributions (normalized the incident electric field amplitude) at 800 nm and $\theta_i = 17°$ angle of incidence. The field maps show strong E field confinement at and around the corners of the pillars. The unit cell boundaries are indicated by white dotted lines in both the top-view and side-view electric-field maps.

Previous measurements of SHG and THG performed on 20 nm- and 70 nm-thick gold unpatterned nanolayers [3] revealed the distinct physical origins of the two processes. SHG in centrosymmetric metals is dominated by surface contributions and is therefore highly sensitive to symmetry breaking at the metal–air interface. As a result, SHG is weakest at normal incidence and increases with angle as symmetry breaking becomes more pronounced. Third harmonic THG, in contrast, is primarily a bulk-mediated process and typically peaks near normal incidence. Our structure has been optimized to support both resonant SHG and THG at moderately oblique incidence angles, large enough to support SHG, but not so large that THG efficiency is compromised. In our previous studies [3, 8], the experimental findings were further supported by a microscopic hydrodynamic model [37, 42-43] for theoretical characterization of the linear and nonlinear optical response of the nanolayer and extraction of the nonlinear parameters, without considering effective values for the second and third order nonlinear susceptibilities. This combined approach provided wavelength-dependent datasets for the dielectric permittivity and third-order nonlinear susceptibility. This model however, proved to be computationally too expensive for 3D simulations. Therefore, in the present study, we incorporate the material parameters and nonlinear source equations derived from the microscopic hydrodynamic framework into a modified COMSOL Multiphysics implementation. This approach enables the 3D numerical modelling of nonlinear SHG and THG emission from our metasurface while retaining the essential microscopic description of the nonlinear optical response.

Although metals lack an intrinsic dipolar second-order susceptibility, they exhibit an effective quadratic nonlinearity arising from several mechanisms, including surface-induced symmetry breaking, magnetic (Lorentz-force) contributions, inner-core electron response, convective nonlinearities, and electron-gas pressure effects [37]. In addition to these effective second-order sources, the third-order nonlinearity of gold also plays a significant role in harmonic generation, as demonstrated in earlier studies of thin metallic films and metal–dielectric photonic crystals [38-39]. A comprehensive treatment that includes all relevant nonlinear contributions, without imposing assumptions about the relative strengths of surface and bulk terms enables accurate prediction of SHG and THG conversion efficiencies, which can reach values on the order of $10^{-7}$ under resonant excitation.

**Second-harmonic generation**

The nonlinear contributions to SHG from nanoscale metals come mostly from free electrons [40-41]. Under the theoretical framework of an expanded Maxwell- hydrodynamic model [37, 42-43] and without any assumptions on the effective surface and volume nonlinearities, the nonlinear equation of motion for the current density of free electrons is written in the time domain as follows:

$$\frac{\partial \mathbf{J}_f}{\partial t}+\gamma_f \mathbf{J}_f=\varepsilon_0\omega_p^2\mathbf{E}+\frac{1}{n_0 e}\left[\mathbf{J}_f(\nabla\cdot\mathbf{J}_f)+(\mathbf{J}_f\cdot\nabla)\mathbf{J}_f\right]-\frac{\mu_0 e}{m^*}\mathbf{J}_f\times\mathbf{H}. \quad (1)$$

Where $\mathbf{J}_f$ is the free-electron current density, $\gamma_f=7.28\times1013\ s^{-1}$ is the damping coefficient associated with gold losses, *e* is the electron charge, $m^*=9.109\times10^{-31}kg$ is the effective electron mass, $\mu_0$ is the vacuum magnetic permeability, $n_0=4.963\times1022\ cm^{-3}$ is the conduction electron density and $\mathbf{H}$ and $\mathbf{E}$ are the local magnetic and electric fields, respectively.

Equation (1) is transformed into the frequency domain by expressing all the variables as superposition of two components, one oscillating at the pump wavelength and the other oscillating at the SH frequency so that we may write the following frequency-domain expression for the current density at the SH frequency [44]:

$$\mathbf{J}_{\mathrm{SH}}=-i\varepsilon_0[\varepsilon_s(\omega_{\mathrm{SH}})-1]\omega_{\mathrm{SH}}\mathbf{E}_{\mathrm{SH}}+\mathbf{J}_{\mathrm{vol}}+\mathbf{J}_{\mathrm{surf}}. \quad (2)$$

The first term on the right-hand side of equation (2) is the linear response at the SH angular frequency $\omega_{SH}$ due to both free and bound electrons, and depends on the experimental permittivity of gold at the SH frequency $\varepsilon_s(\omega_{SH})$. The second ($\mathbf{J}_{vol}$) and third ($\mathbf{J}_{\mathrm{surf}}$) terms are bulk and surface nonlinear contributions due to free electrons. For vacuum–metal interfaces, taking into account absorption losses in the metal and the effects of a generic dielectric with permittivity $\varepsilon B$ in contact with metal, these terms can be writtes as[37]:

$$\mathbf{J}_{\mathrm{vol}}=i\omega_{\mathrm{SH}}\varepsilon_0\frac{e}{2m^*\omega_{\mathrm{FF}}^2}\frac{\varepsilon_{s,r}(\omega_{\mathrm{FF}})-1}{2}\beta\left[(\alpha-1)(\mathbf{E}_{\mathrm{FF}}\cdot\nabla)\mathbf{E}_{\mathrm{FF}}+\frac{1}{2}\nabla(\mathbf{E}_{\mathrm{FF}}\cdot\mathbf{E}_{\mathrm{FF}})\right]. \quad (3)$$

$$\hat{\mathbf{t}}\cdot\mathbf{J}_{\mathrm{surf}}=-i\omega_{\mathrm{SH}}\varepsilon_0\chi_{\parallel}^{(2)}E_{\mathrm{FF},\parallel}E_{\mathrm{FF},\perp}, \quad (4.1)$$

$$\mathbf{n}\cdot\mathbf{J}_{\mathrm{surf}}=-i\omega_{\mathrm{SH}}\varepsilon_0\chi_{\perp}^{(2)}E_{\mathrm{FF},\perp}^2. \quad (4.2)$$

In expressions (3) and (4) $\omega_{FF}$ is the pump angular frequency, $\varepsilon_{s,f}(\omega_{FF})=-\frac{e^2 n_0}{m^*(\omega_{FF}^2+i\omega_{FF}\gamma_f)}$ is the free-electron permittivity, $\alpha=\frac{\omega_{FF}}{(\omega_{FF}+i\gamma_f)}$, $\beta=\frac{\omega_{SH}}{(\omega_{SH}+i\gamma_f)}$ and $\boldsymbol{E}_{FF}$ is the electric field at the pump frequency; $\hat{\mathbf{t}}$ and $\mathbf{n}$ define a local, boundary coordinate system and represent unit vectors pointing in the directions tangential and outward normal to the metallic surfaces, respectively; $\boldsymbol{E}_{FF,\parallel}$, and $\boldsymbol{E}_{FF,\perp}$ are the magnitudes of the tangential and normal electric field components at the pump frequency in this local coordinate system. The surface, quadratic nonlinear susceptibilities are then given by:

$$\chi_{\parallel}^{(2)} = -\alpha\beta \frac{e}{2m^*\omega_{\mathrm{FF}}^2}\left[\frac{\varepsilon_s(\varepsilon_B - 1) + \varepsilon_B\left(1-\varepsilon_{s,f}(\omega_{\mathrm{FF}})\right)}{\varepsilon_B}\right] \tag{5.1}$$

$$\chi_{\perp}^{(2)} = -\alpha\beta \frac{e}{2m^*\omega_{\mathrm{FF}}^2}\left[\frac{3\varepsilon_s^2(\varepsilon_B - 1) + \varepsilon_B\varepsilon_s\left(\varepsilon_B - \varepsilon_{s,f}(\omega_{\mathrm{FF}})\right) + 3\varepsilon_B^2\left(1-\varepsilon_{s,f}(\omega_{\mathrm{FF}})\right)}{4\varepsilon_B^2}\right] \tag{5.2}$$

Where $\varepsilon_B$ assumes the value $\varepsilon_c$ or $\varepsilon_h$ depending on whether the inner or outer interface of the shell is considered for $\chi_{\parallel}^{(2)}$ and $\chi_{\perp}^{(2)}$ calculations. In 3D simulation platform, $E_{FF} = E_x\hat{i} + E_y\hat{j} + E_z\hat{k}$, thus Eqs. (3), (4) and (5) are also deconstructed into their corresponding x, y and z components, to be introduced into the model; $\mathbf{J}_{\mathrm{vol}}$ is introduced as the nonlinear polarization term in the solution of Maxwell's equations, whereas $\hat{\mathbf{t}} \cdot \mathbf{J}_{\mathrm{surf}}$ and $\mathbf{n} \cdot \mathbf{J}_{\mathrm{surf}}$ are defined as surface currents at every metal-air interface.

**Third-harmonic generation**

THG, by contrast, is largely a bulk-mediated process. To simulate bulk THG from gold, structured or unstructured, the nonlinear polarization term in introduced into the Maxwell's equation of EM beam propagation as follows:

$$\nabla \times \mu_r^{-1}(\nabla \times \mathbf{E}) - k_0^2(\varepsilon_r - \frac{j\sigma}{\omega\varepsilon_0})\mathbf{E} = \omega^2\mu_0\sum_i \mathbf{P}_i \tag{6}$$

The following third order nonlinear effects were introduced via the nonlinear polarization term in the abovementioned EM wave equation.

1. The Kerr effect (added as a nonlinear polarization term at the pump frequency):

$$\mathbf{P}_\omega = \chi_\omega^{(1)}\mathbf{E}_\omega + 3\varepsilon_0\chi_\omega^{(3)}\left|\mathbf{E}_\omega\right|^2\mathbf{E}_\omega \tag{7.1}$$

2. Third harmonic generation (added as a nonlinear polarization term at the third harmonic frequency):

$$\mathbf{P}_{3\omega} = \chi_{3\omega}^{(1)}\mathbf{E}_{3\omega} + \chi_{3\omega}^{(3)}(\mathbf{E}_\omega.\mathbf{E}_\omega)\mathbf{E}_\omega \tag{7.2}$$

In principle, the third-order susceptibility $\chi^{(3)}$ is a rank-4 tensor, containing up to 81 components. However, for isotropic and centrosymmetric media such as gold, due to inversion symmetry, the tensor structure gets drastically simplified. Gold has a centrosymmetric crystal lattice and behaves as an optically isotropic metal at optical frequencies. Under these symmetry constraints, all off-diagonal tensor elements of the $\chi^{(3)}$ tensor vanish, and the remaining diagonal elements become equal. As a result, $\chi^{(3)}$ gets reduced to a single independent scalar parameter. The hydrodynamic description provides a microscopic framework for understanding the nonlinear response of materials at the nanoscale. In metallic nanostructures, the interplay between spatially confined resonant EM fields and nonlinear electron-response mechanisms determines the efficiency and spatial distribution of the generated harmonic fields. As shown in Fig. 1(d) and (e), the local electric-field intensity is enhanced by a factor of approximately 38. Consequently, an incident pump intensity of a few GW/cm² can produce local intensities of several TW/cm² near the nanopillar corners. Under these strongly confined field conditions, an accurate description of the nonlinear optical response requires consideration of the interplay between free- and bound-electron contributions.

In Fig. 2(a) we show schematics of a unit cell of the gold nano-pillars as implemented in COMSOL , where we indicate the surfaces to implement the Eqs. (4.1) and (4.2). In Fig. 2(b) we indicate the bulk of the material to implement the Eqs. (3), (7.1) and (7.2). In Fig. 2(c) and (d) we plot the theoretically predicted spectral response of the resonant SHG and THG from the gold nanopillars metasurface respectively, along with the same from plain layer of 150 nm thick gold (purple curve). The structure and the plain gold are both illuminated by pump of 800 nm at $\theta_B = 17^\circ$ angle of incidence. The theoretical predictions suggest that resonant SHG produces a peak conversion efficiency of $\eta_{SH} = \frac{P_{SH}}{P_{FF}} = 7.7676 * 10^{-8}$ (Fig. 2(c)), corresponding to an enhancement factor of ~651, with respect to the SHG from plain gold, under the exact same pump conditions. The resonant THG reaches a peak conversion efficiency of $\eta_{TH} = \frac{P_{TH}}{P_{FF}} = 8.78*10^{-7}$ (Fig. 2(d)), corresponding to an enhancement factor of ~1307.

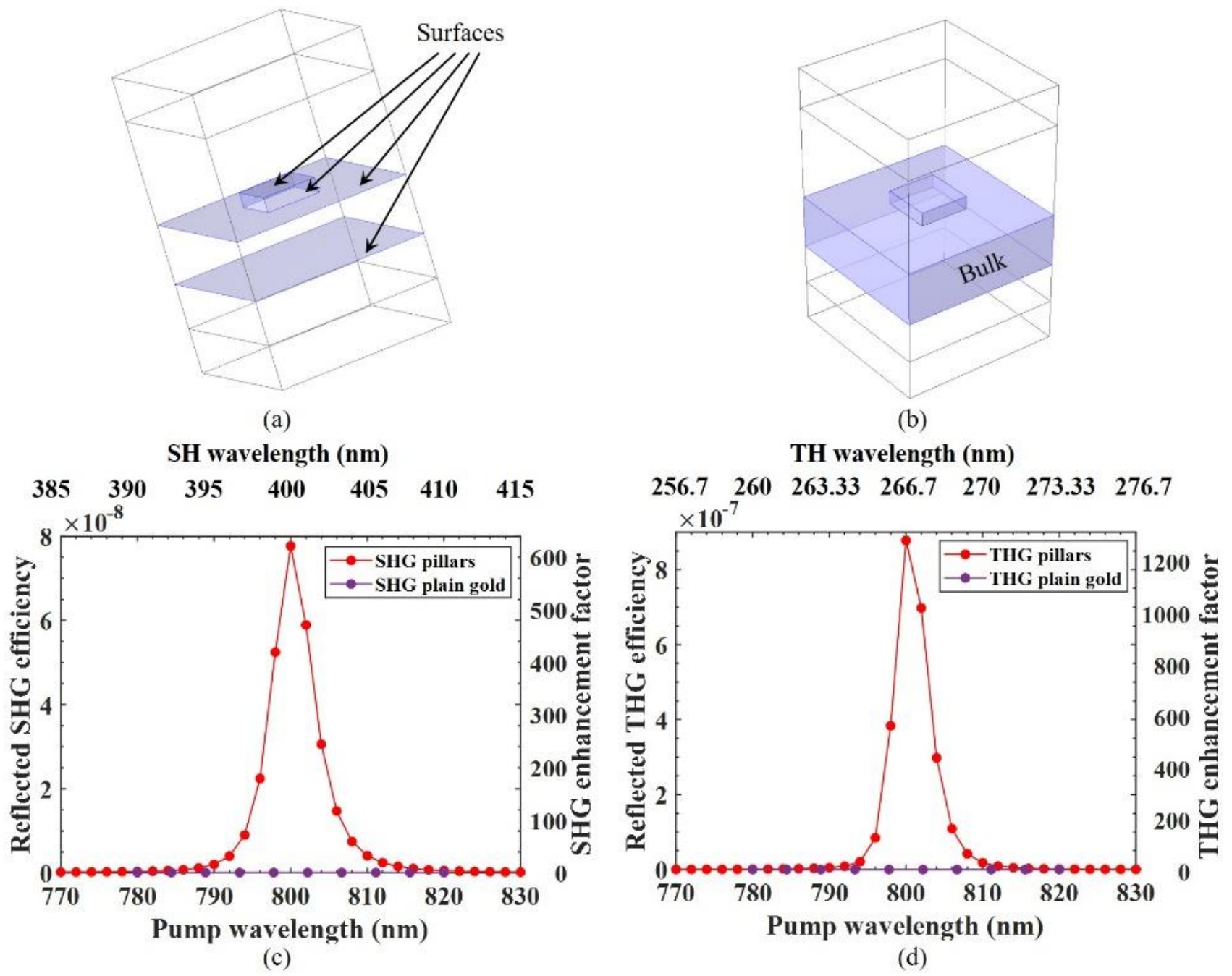

Fig. 2(a) Surfaces selected to define free electron contribution to SHG (b) Bulk selected to define bulk SHG and THG polarizations. (c) Theoretically estimated SHG conversion efficiency for 17° angle of incidence (d) Theoretically estimated THG conversion efficiency for 17° angle of incidence. In each case, the left y-axis represents the pump to SHG/THG conversion efficiency and the right y-axis represents the enhancement factor of the SHG/THG efficiency compared to that from the unstructured metal (purple curve).

## 3. Fabrication and Experimental characterization

### 3.1. Fabrication

The fabrication process of the gold nanopillar metasurfaces is schematically illustrated in Fig. 3(a). The process began with a crystalline silicon (c-Si) wafer, which was subsequently diced into $15 \times 15\text{mm}^2$ samples using a Disco DAD3220 dicing saw system. A 150 nm thick gold layer was then deposited onto the c-Si substrate using an electron-beam evaporation system (CHA Industries Solution E-Beam) at a deposition rate of 5 Å/s, forming the base metallic layer of the structure. Following metal deposition, the samples were prepared for electron-beam lithography (EBL). A 50 nm thick layer of PMMA 950A2 electron-beam resist was spin-coated onto the sample surface, followed by a prebake at 180°C to remove residual solvent and solidify the resist layer. Although gold is electrically conductive, a 40 nm thick conductive discharge layer (DisCharge $H_2O$ x2, DisChem) was additionally spin-coated to further suppress charging effects during EBL exposure. Pattern definition was subsequently carried out using an Elionix ELS-7500 EX electron-beam lithography system, followed by resist development to create the nanopillar mask pattern. After completion of the lithographic process, a second 50 nm thick gold layer was deposited using the same electron-beam evaporation system. Finally, the sample was immersed in Microposit Remover 1165 (NMP) for approximately 12 hours to perform the lift-off process, resulting in the final periodic gold nanopillar metasurface structures. Fig. 3(b) shows a top view scanning electron microscope (SEM) image of the fabricated sample, indicating near perfect square lattice with the following geometric parameters: the period (p=605.3 nm in one direction and p=611.9 nm in the other direction), and the width of the pillars (d=185.2 nm).

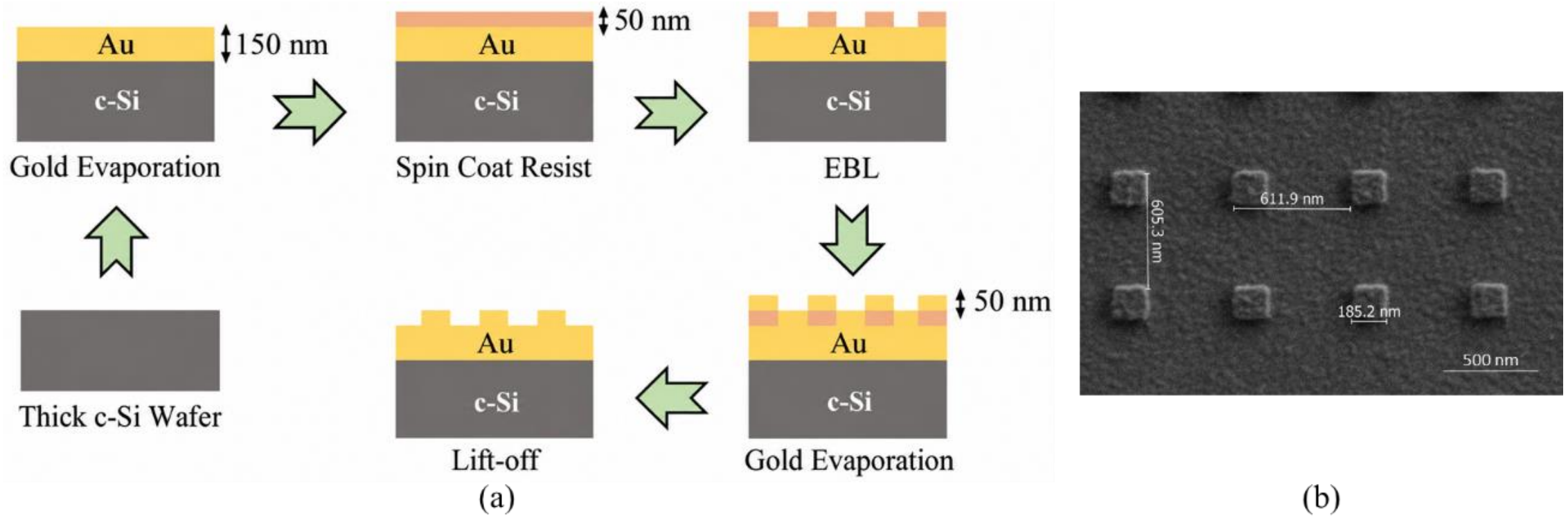

Figure 3. (a) Schematic illustration of the fabrication process for the gold nanopillar metasurface. A 150 nm thick gold film is first deposited onto a crystalline silicon substrate by gold evaporation. A 50 nm thick electron-beam resist layer is then spin-coated on the sample, followed by electron-beam lithography to define the nanopillar pattern. Subsequently, an additional 50 nm gold layer is evaporated, and a lift-off process is performed to obtain the final periodic gold nanopillar array on the c-Si substrate. (b) SEM image top view of the sample indicating the periodicity and width of the pillars.

### 3.2. Linear characterization

To characterize The linear optical response of the metasurface has been characterized using  a supercontinuum CW laser source (FYLA SCT 1000) and collecting the reflected light using a high-resolution spectrometer. By tuning the angle of incidence from 12° to 20° we recorded the reflected spectra from the patterned metasurface region as well as from the adjacent unpatterned gold film.

The linear characterization serves to identify the plasmonic resonances that govern the subsequent nonlinear frequency-conversion processes. Owing to the thickness of the underlying gold, transmission through the sample is negligible over the investigated spectral range, a minima in the reflectance spectra correspond to enhanced optical absorption associated with the excitation of plasmonic resonances.

In Fig. 4(a) we display an image of the sample showing the patterned and unpatterned sections of the gold. Figure 4(b) shows the experimentally retrieved angular plot of the reflected spectra from the periodic pillars, when normalized to the reflected spectra from the unpatterned part of the gold. The plot depicts the tunable resonance, as the angle of incidence of the pump beam is increased, the spectral position of the resonance is red shifted, as predicted previously by numerical simulations. Experimentally retrieved depth of resonance is not as deep as predicted by simulations in Fig. 1(d). This discrepancy may be attributed to slight deviations from the ideal design parameters assumed in the simulations, such as rounded corners of the pillars instead of perfectly 90° corners, non-vertical walls, slight differences in nano pillars spacing and depth within the laser spot, as assumed in our simulations. These parameters also affect the strength of light localization at the corners of the gold pillars, and we can safely assume the magnitude of E field localization in reality is lower than as predicted in Figs. 1(d) and 1(e). We find three angles with resonances located near 800 nm, i.e. $\lambda_A = 795nm$ for $\theta_A = 16^\circ$, $\lambda_B = 806nm$ for $\theta_B = 17^\circ$, and $\lambda_C = 815nm$ for $\theta_C = 18^\circ$. In these three cases, the SHG and THG signals fall within the transmission bandwidth of the specific narrow bandpass filters used in our nonlinear detection systems, as described in Fig. 5(a) and Supplementary info Fig. S2. Our following nonlinear measurements and analysis will be carried out in these three pump illumination conditions.

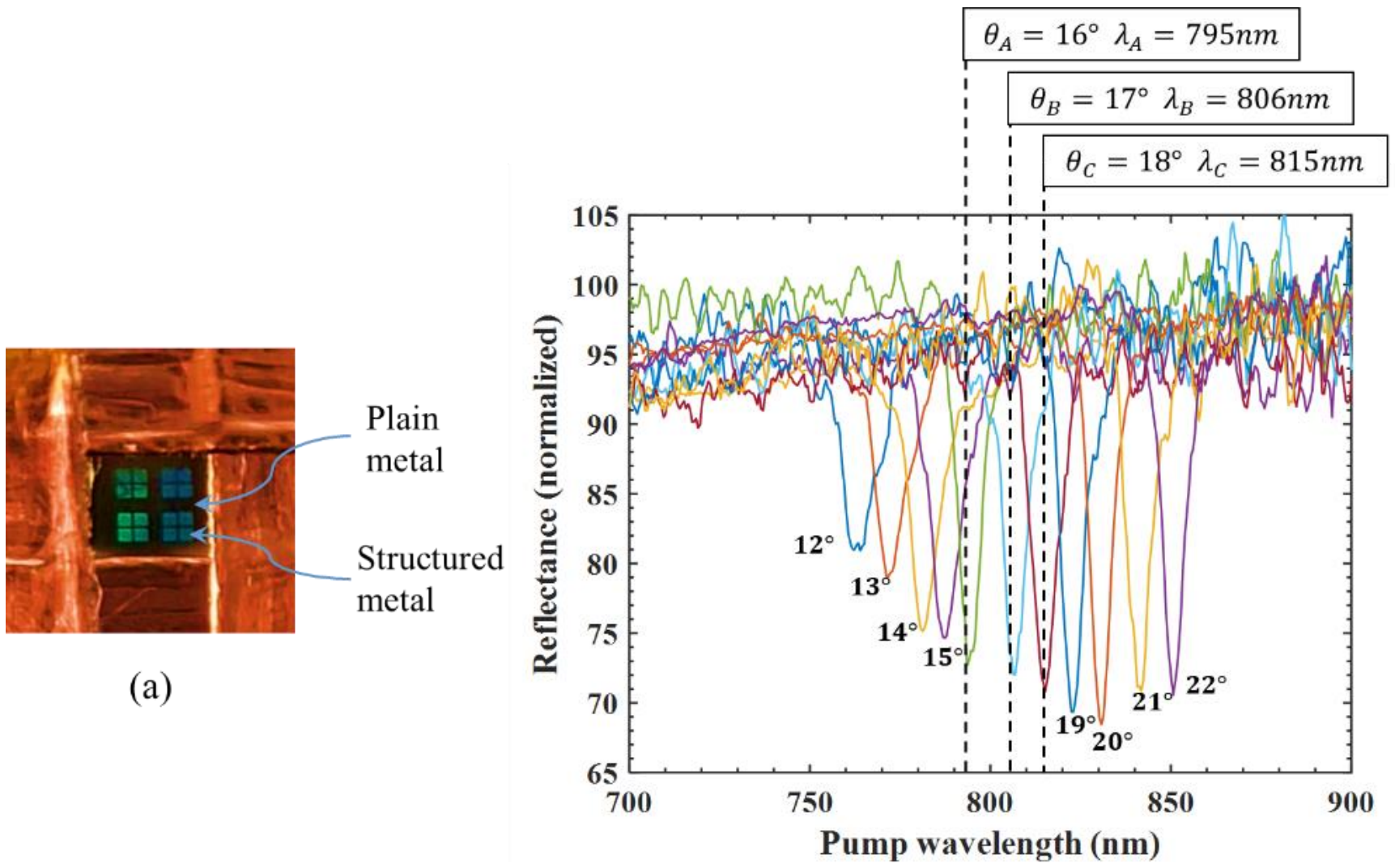

(a) (b)

Fig. 4(a) Image of the sample showing the patterned and unpatterned sections of the gold. (b)Experimentally retrieved linear reflectance spectra with varying angle of incidence for TM polarized broadband pump illumination; reflectance from the periodic gold pillars is normalized (%) to the reflectance from the unpatterned plain gold. The spectral position of the resonance is indicated for the three target resonances at $\theta_A = 16^\circ$, $\theta_B = 17^\circ$, and $\theta_C = 18^\circ$.

### 3.3. Nonlinear characterization

For nonlinear characterization we constructed the optical setup illustrated in Fig.5 (a). The arrangement enables quantitative measurements of the power conversion efficiency from the incident pump to the THG frequency, defined as $\eta_{SH/TH} = \frac{P_{SH/TH}}{P_{Pump}}$, with detectable values ranging from $10^{-6}$ down to $10^{-12}$ level. For wavelength-dependent nonlinear measurement, a separate fiber-amplified femtosecond laser is used as the pump, which is equipped with an optical parametric amplifier (CARBIDE+OPA from Light Conversion) tunable from 650 nm to 2500 nm. The system delivers pulses of approximately of 150 fs duration at a 60 kHz repetition rate, with pulse energy up to 6 µJ and a maximum average power of 400 mW. The output polarization (TE or TM) is controlled using a calcite Glan-laser polarizer followed by a continuously rotating half-wave plate. A near-IR long-pass filter suppresses any spurious nonlinear signals generated by other optical components along the beam path. The pump beam is focused onto the sample using a plano-convex lens of f=200 mm, yielding peak intensities of the order of ~10-12 GW/cm$^2$. Each sample has a square cross-sectional area of $250\mu m \times 250\mu m$ patterned region, sufficiently large to accommodate the focused beam, which has a diameter of approximately 97 µm (calculated at $e^{-2}$intensity) at normal or near normal incidence. During wavelength sweeps, the pump wavelength was changed manually. For each wavelength, the laser energy is tuned to compensate for the OPA's wavelength dependent conversion efficiency. In addition, wavelength specific responsivity values were applied to the detection system, and a correction factor was used to account for the wavelength-dependent transmission (or reflection) of all optical elements in the beam path. The samples are mounted on an XYZ translation stage attached to a motorized rotation platform, allowing precise control of the incidence angle with a resolution of 0.2°.

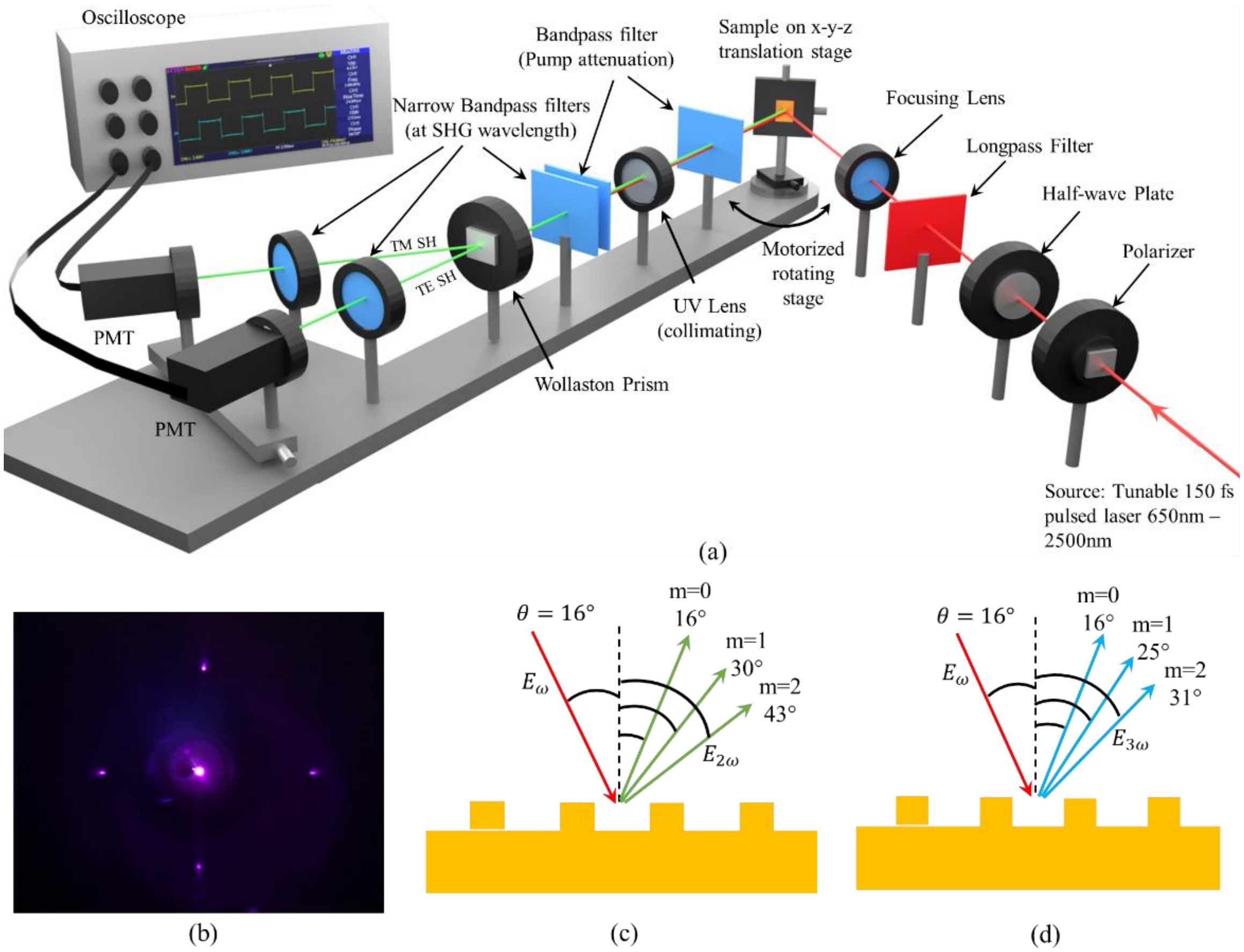

Figure 5. (a) Schematic illustration of the experimental set-up used to measure SHG in reflection from the gold nanopillars samples, as a function of angle of incidence, polarization, and pump wavelength. (b) Experimental verification of the diffraction pattern of the SHG wavelength [8] (c) Schematic representation of the diffracted SHG. (d) Schematic representation of the diffracted THG.

Fig. 5 (a) also illustrates the SH detection arm. Right after the sample, a broadband bandpass filter (Thorlabs, FGB37-A) is placed that attenuates the pump field that has passed through the sample, followed by an UV grade fused silica lens, to collimate the beam. However, this bandpass filter is not enough to eliminate all the pump radiation, and we place another FGB37-A broadband bandpass filter after the collimating lens, along with two narrow bandpass spectral filters that have a 20 nm bandpass transmission around the SH frequency (Thorlabs, FBH400-40), one of which is placed just in front of the detector. By choosing these filters, the total optical density exceeds the harmonic generation efficiency, ensuring that only SH radiation arrives at the detector. The collimator lens placed after the sample provides an angular admission range of roughly 10°. A Wollaston polarizer separates the TM and TE polarization components of the SH signal, allowing simultaneous measurements of both polarizations. We expect values of SH efficiency of the order of $10^{-9}$ or $10^{-10}$, requiring a very sensitive detection system. Light is collected with a photomultiplier (Hamamatsu, H10722-113 and H10722-04) where the detected signal is strongly amplified. Finally, the PMTs are connected to their controllers and to an oscilloscope to record the generated harmonic signals. The entire detection system is mounted on a rigid platform, which in turn is placed on a rotating tail to allow the measurement of either transmittance or reflectance.

For detection of the TH signals in the deep-UV spectral range, we used optical components and filters specifically optimized for high transmission and performance at the target wavelengths. (as detailed in [45] and demonstrated in Supplementary info Fig. S2). A detailed alignment and comprehensive calibration procedure, accounting for the transmission characteristics of every optical element placed after the sample i.e. filters, lenses, and mirrors, as well as the wavelength dependent responsivity of the PMT, allows us to accurately estimate the conversion efficiencies as the ratio of the SH/TH energy to the energy of the incident pump beam. The entire detection arm, which is the most critical part of the setup, is mounted on a rotary platform that enables a precise setting of the detection angle for reflection measurements. We also verified that the high average excitation levels did not induce cumulative thermal effects or long-term degradation and optical inspection of the samples after the experiments revealed no visible damage. For measurements in other wavelength ranges, the suitable optical elements are used and the detection is recalibrated.

Given the grating periodicity of ~610 nm, the pump beam does not generate additional diffraction orders, while its SH (around 400 nm) and TH (around 266 nm) will produce several diffraction orders each. Fig. 5(b) presents the corresponding 2D periodic diffraction pattern, obtained when the structure is artificially illuminated by a low power CW laser source of wavelength 400 nm i.e. the second harmonic wavelength. Fig. 5(c) shows the zeroth and first two diffraction orders of the generated SH, which were experimentally detected as follows: the angle of incidence of the pump beam was fixed at $\theta_i = 16°$ while the detection arm was moved horizontally in the plane of incidence, performing a full angular scan (+/- 90°). Fig. 5(d) is a schematic of the THG diffraction orders, detected in the abovementioned way. In this way we retrieved the relative intensity contributions of the different diffraction orders at first two diffraction orders of the generated SH/TH, on each side of the transmitted zero order. Limitations in our experimental setup allowed us to measure only diffraction orders appearing at positive angles on the side of the reflecting beam. However, due to the 2D symmetric periodicity of the structure, the positive diffraction orders are generate not only horizontally, but also vertically, which were also taken in to account. The relative intensity of the different diffraction orders were recorded, and their sum helps us retrieve the total conversion efficiency to correctly corroborate our theoretical predictions, as we simulate the total transmitted light passing through and integrated over the area of transmission.

**Second-harmonic generation**

The first set of nonlinear measurements are carried out to retrieve the resonant SHG from the periodic gold pillars at 16°, 17°, and 18° angles of incidence, the TM polarized pump wavelength is tuned around the wavelength of resonance in each case. The individual linear resonances at these angles are shown in Fig. 6(a), in each case we plot the reflectance from the periodic gold pillars structure, normalized to the reflectance obtained from the plain sputtered gold region of the sample.

The corresponding experimental measurements of total SHG at $\theta_A = 16^\circ$, $\theta_B = 17^\circ$, and $\theta_C = 18^\circ$ angles of incidence, as a sum of all diffracted orders, are represented in Fig. 6(b). The experimental points correspond to SHG signals retrieved every 1 nm, while the pump wavelength is varied around the spectral position of the linear resonance, 1 nm being the spectral resolution of the source laser system. For each curve, the left y-axis represents the experimental SHG/THG efficiency and the right y-axis represents the enhancement factor of the experimental SHG/THG efficiency compared to that from the unstructured metal, under the same laboratory environment and illumination. Thus when the structure is illuminated by pump at 16° angle of incidence, the spectral measurement of the resonant SHG shows a peak conversion efficiency of $\eta_{TH} = \frac{P_{SH}}{P_{FF}} = 2.09958*10^{-8}$, corresponding to an enhancement factor of ~258. For ease of understanding, we record the spectral position of the resonances, resonant SHG conversion efficiencies and the corresponding SHG enhancement factors in Table1. Because of the 2D symmetry of the structure, similar angular response and nonlinear conversion efficiencies are expected for TE polarized pump illumination.

| Angle of Incidence | $\theta_A = 16^\circ$ | $\theta_B = 17^\circ$ | $\theta_C = 18^\circ$ |
|---|---|---|---|
| $\lambda_{Res}$ at Pump | $\lambda_A = 795nm$ | $\lambda_B = 806nm$ | $\lambda_C = 815nm$ |
| $\lambda_{Res}$ at SHG | 395 | 400.5 | 406 |
| Pump to SHG conversion efficiency (peak) | $2.09958*10^{-8}$ | $2.35165*10^{-8}$ | $2.61595*10^{-8}$ |
| SHG enhancement (peak) w.r.t. plain gold | 258 | 284 | 318.6 |

Table 1: Angle-dependent SHG response of the gold nanopillars metasurface: Pump and SHG wavelengths at resonance, the corresponding peak SHG conversion efficiencies, and enhancement factors relative to a plain gold film.

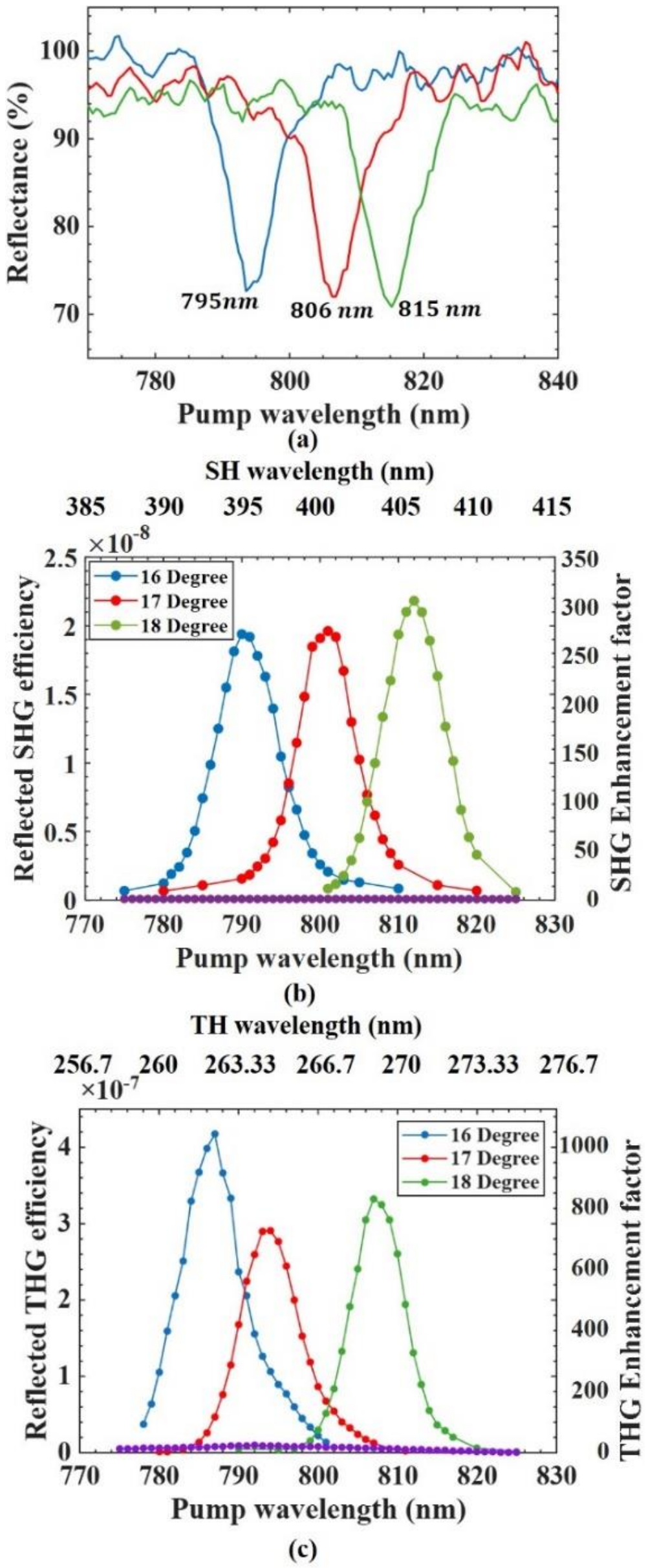

Fig. 6(a) Spectral position of the three target resonances at $\theta_A = 16^\circ$, $\theta_B = 17^\circ$, and $\theta_C = 18^\circ$ angles of incidence (b) Resonant SHG spectral peaks corresponding to each angle of incidence (c) Resonant THG spectral peaks corresponding to each angle of incidence. In (b) and (c), the left y-axis represents the experimentally retrieved pump to SHG/THG conversion efficiency and the right y-axis represents the enhancement factor of the experimental SHG/THG efficiency compared to that from the unstructured metal (purple curve), under the same laboratory environment and illumination.

**Third-harmonic generation**

The second set of nonlinear measurements are carried out to retrieve the resonant THG from the periodic gold pillars at $\theta_A = 16^\circ$, $\theta_B = 17^\circ$, and $\theta_C = 18^\circ$ angles of incidence, the pump wavelength is tuned around the wavelength of resonance in each case. The corresponding experimental measurements of total THG at $\theta_A = 16^\circ$, $\theta_B = 17^\circ$, and $\theta_C = 18^\circ$ angles of incidence, as a sum of all diffracted orders, are represented in Fig. 6(c). When the structure is illuminated by pump at $\theta_A = 16^\circ$ angle of incidence, the spectral measurement of the resonant THG shows a peak conversion efficiency of $\eta_{TH} = \frac{P_{TH}}{P_{FF}} = 4.17152*10^{-7}$, corresponding to an enhancement factor of ~1060. Once again, we record the spectral positions of the resonances, resonant THG conversion efficiencies and the corresponding THG enhancement factors in Table2.

| Angle of Incidence | $\theta_A = 16^\circ$ | $\theta_B = 17^\circ$ | $\theta_C = 18^\circ$ |
|---|---|---|---|

| $\lambda_{\text{Res}}$ at pump | $\lambda_A = 795nm$ | $\lambda_B = 806nm$ | $\lambda_C = 815nm$ |
|---|---|---|---|
| $\lambda_{\text{Res}}$ at THG | 262.33 | 264.33 | 269 |
| Pump to THG conversion efficiency (peak) | $4.17152 * 10^{-7}$ | $2.9107 * 10^{-7}$ | $3.3253 * 10^{-7}$ |
| THG enhancement (peak) w.r.t. plain gold | 1060 | 705 | 945 |

Table 2: Angle-dependent THG response of the gold nanopillars metasurface: Pump and THG wavelengths at resonance, the corresponding peak THG conversion efficiencies, and enhancement factors relative to a plain gold film.

The experimentally measured resonant SHG and THG peaks are blue-shifted by a few nanometers in both cases. To investigate the origin of these shifts, THG measurements were repeated on the same day under identical experimental conditions, using three different pump intensities. The resonance peak remained at the same spectral position for all three intensities (see Supplementary info Fig. S3), suggesting that nonlinear or cumulative thermal effects are not responsible for the observed blue shift. We therefore attribute the small blue shifts observed in the measured peaks to minor experimental differences between the setups used for linear and nonlinear measurements, i.e. pump beam size, different detectors and different rotating stages.

While the estimates of experimental SH/TH conversion efficiencies and the corresponding enhancement factors are limited by absorption and losses in real laboratory environment, our theoretical predictions suggest that they can potentially be better (Fig. 2© and (d)). This discrepancy may once again be attributed to slight deviations of the fabricated structures from the ideal geometrical parameters assumed in the simulations. In particular, fabrication imperfections such as rounded pillar corners instead of perfectly sharp 90° edges, non-vertical sidewalls, and small variations in nanopillar spacing and height across the illuminated area can modify the local EM field distribution. In reality, the amplitude of E field localization at the corners of the fabricated samples is lower than as predicted in Figs. 1(d) and 1(e), thus justifying experimental retrieval of slightly lower SHG/THG. Nevertheless, the experimentally retrieved enhancement factors are unprecedented in comparison to previously reported values for purely plasmonic gold metasurface platforms.

## 4. Conclusion

In this work, we have designed, fabricated, and experimentally characterized a resonant plasmonic metasurface composed of 2D periodic gold nanopillars, engineered for enhanced SHG and THG under femtosecond lase pulse excitation. The results highlight the critical role of nanoscale resonant EM field confinement and angle-tunable plasmonic resonances in boosting ultrafast nonlinear optical interactions and enabling frequency upconversion into the UV and deep-UV spectral regions. By accurately detecting and calibrating the nonlinear conversion efficiencies, we quantified the resonant enhancement achieved by the metasurface relative to an unstructured gold film, demonstrating two orders of magnitude enhancement in SHG and three orders of magnitude enhancement in THG. To the best of our knowledge, these represent the highest orders of enhancement reported for SHG and THG in this type of purely metallic metasurface. The substantial THG enhancement further demonstrates the potential of resonant metallic nanostructures to promote higher-order nonlinear optical processes and extend ultrafast frequency conversion into the deep-UV, while also highlighting the successful detection and quantitative characterization of extremely weak nonlinear signals in this spectral region.

Our results highlight that metasurface geometry i.e. periodicity, feature size, and overall symmetry play a decisive role in determining the strength, spectral position, and polarization response of plasmonic resonances. These geometric degrees of freedom directly influence both the local field distribution and the effective nonlinear susceptibility, underscoring the need for a comprehensive understanding of the physical parameters governing plasmonic behavior, including bulk and surface nonlinearities. Beyond frequency conversion, the demonstrated platform is inherently versatile. The combination of strong field confinement, angle-tunable resonances, deep-UV harmonic generation, and accurate nonlinear modeling positions this metasurface as a promising foundation for several high-value photonic technologies, including compact UV sources, on-chip frequency converters, ultrasensitive detectors, and advanced spectroscopic tools.

The demonstrated metasurface platform illustrates how nanoscale architectural control enables precise manipulation of EM hotspots, leading to substantial enhancement of light–matter interactions and improve harmonic-generation efficiency. As nano fabrication techniques continue to advance, the ability to tailor resonant responses at the nanoscale will further expand the design space for high-performance nonlinear photonic devices. Such capabilities position plasmonic metasurfaces as promising candidates for a broad range of linear and nonlinear applications, including compact UV sources, on-chip frequency converters, ultrasensitive detectors, and advanced spectroscopic tools.

## Acknowledgements

SM, JT, and CC acknowledge the Spanish Agencia Estatal de Investigación from Ministerio de Ciencia, Investigación y Universidades (project PID2023-148620NB-I00 granted by MICIU/AEI/10.13039/501100011033 and FEDER, UE) and AGAUR (Project 2025SGR00171). S. M. acknowledges predoctoral grant FPI-UPC 2022, funded by Universitat Politècnica de Catalunya (Government of Catalunya). NML acknowledges the US Office of Naval Research award N00014-25-1-2402 and the US Army Research Office award W911NF-26-1-A281.